\documentclass[twocolumn,showpacs,preprintnumbers,amsmath,amssymb,showkeys]{revtex4}

\usepackage[dvips]{graphicx}
\usepackage{dcolumn}
\usepackage{bm}

\begin{document}


\title{Solvable Model of Spiral Wave Chimeras}

\author{Erik A. Martens}
\affiliation{Max Planck Institute for Dynamics and Self-Organization, 37073 G\"{o}ttingen, Germany}

\author{Carlo R. Laing}
\affiliation{ IIMS, Massey University, Private Bag 102-904 NSMC, Auckland, New Zealand}

\author{Steven H. Strogatz}
\affiliation{Department of Mathematics, Cornell University, Ithaca, NY 14853, USA}

\date{\today}

\begin{abstract}
Spiral waves are ubiquitous in two-dimensional systems of chemical or biological oscillators coupled locally by diffusion.  At the center of such spirals is a phase singularity, a topological defect where the oscillator amplitude drops to zero.  But if the coupling is nonlocal, a new kind of spiral can occur, with a circular core consisting of desynchronized oscillators running at full amplitude.  Here we provide the first analytical description of such a spiral wave chimera, and use perturbation theory to calculate its rotation speed and the size of its incoherent core. 
\end{abstract}

                             

\maketitle

Large systems of coupled limit-cycle oscillators have been used to model diverse phenomena 
in physics, chemistry, and biology~\cite{winfreebook80,pikovskybook03}.  Examples range from spiral waves in heart muscle~\cite{pauerm94} and 
brain tissue~\cite{horoph97}, 
to synchronization of nerve 
cells~\cite{chokop00}, pendulums~\cite{pantaleone02}, fireflies~\cite{buck88}, 
and Josephson junctions~\cite{wiesenfeld96}.

Most of the early theoretical work on these systems assumed either
local coupling (through nearest-neighbor or diffusive interactions) or global coupling 
(through infinite-range interactions, corresponding to a mean-field approximation).   
In the past few years, however, several researchers have begun to explore other types of connectivity.  
One line of research
investigates what happens if the coupling is neither local nor global but 
somewhere in between, as occurs naturally in certain neural, chemical, and biochemical 
systems~\cite{laicho01, kurbat02, kura03, tankur03}. 

Around 2002, Kuramoto and his colleagues discovered a spatiotemporal pattern that appears to be unique to such nonlocally coupled 
systems~\cite{kurbat02, kura03}.  They showed numerically that systems of identical 
oscillators with symmetrical coupling could self-organize into a state 
in which some oscillators were mutually synchronized while others remained desynchronized.  
An exact solution for this ``chimera state'' was later obtained for phase oscillators 
arranged on a ring \cite{abrstr04,abrstr06}, and its dynamics and bifurcations 
were further clarified by studying two interacting populations of 
oscillators \cite{abrmir08,pikovskyPRL08,lai09A}. 

The most surprising result emerged from simulations of two-dimensional 
arrays \cite{shikur04, lai09B,shikur03}.  Kuramoto and coworkers \cite{shikur03,shikur04,kurshi06} 
showed that two-dimensional chimeras are spiral waves with a 
phase-randomized core of desynchronized oscillators surrounded by phase-locked 
oscillators in the spiral arms.  Nothing like this had ever been seen before; in ordinary reaction-diffusion systems, the smoothing effects of diffusion would prevent the spatial discontinuities implied by an incoherent core.

In this Letter we present the first analytical solution for a spiral wave chimera.  Our analysis predicts the existence of the incoherent core and yields a simple formula for its radius as well as for the rotation rate of the spiral arms.    

The model we consider is 
\begin{equation}
    \frac{\partial \phi({\bf x},t)}{\partial t}=
\omega-\int_{\mathbb{R}^2}\!\! G(|{\bf x-x'}|)\sin{[\phi({\bf x},t)-\phi({\bf x'},t)+\alpha]}d{\bf x'} \label{eq:dphi}
\end{equation}
where $\phi({\bf x},t)$ is the phase of the oscillator at position ${\bf x} \in \mathbb{R}^2$ 
at time $t$, $\omega$ is its natural frequency, and $\alpha$ is a phase lag.  The kernel $G$ is a 
normalized Gaussian:
\begin{equation}
   G(|{\bf x-x'}|)=\frac{e^{-|{\bf x-x'}|^2}}{\pi} .  \label{eq:G}
\end{equation}
Note that the width of $G$ determines a length scale for the
system. Varying $\alpha$ can be thought of as choosing different first harmonics of a general
coupling function~\cite{koerm08}.  In contrast, $\omega$ is dynamically irrelevant, and can be set to zero without loss of generality.   

\begin{figure}
	\includegraphics[width=3.2in]{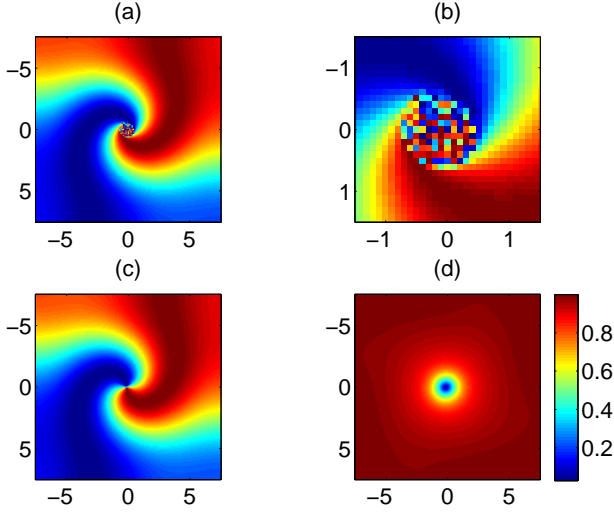}
	\caption{\label{fig:spi1} Snapshot of a spiral wave chimera on a square domain of side length $15$, 
	using an array of $151\times 151$ oscillators. Parameter values: $\alpha=0.15 \pi$, $\omega=0$.  Colors schematically indicate oscillation phases in panels (a)-(c), whereas in panel (d) they depict the amplitude $R(\mathbf{x})$ of the local order parameter according to the numerical values shown next to the color bar.
	(a): Spatial variation of $\sin{\phi}$. 
	(b): Close-up of the incoherent core in panel (a). 
	(c):  $\sin{(\Theta(\mathbf{x}))}$ 
(phase of order parameter)
	(d): $R(\mathbf{x})$ (amplitude of order parameter). The amplitude drops to zero at the center of the core, consistent with the randomized phases of the oscillators there.}   
\end{figure}

Figure~\ref{fig:spi1}(a) shows a spiral wave chimera for Eqs.~(\ref{eq:dphi}), (\ref{eq:G}), computed numerically.  Notice the phase-randomized core at its center, as highlighted in Fig.~\ref{fig:spi1}(b).  The spatial structure of this pattern is governed by an integral equation, originally derived by Shima and Kuramoto~\cite{shikur04} via the following self-consistency argument.  Define a local mean field (a complex order parameter) by
\begin{equation}
    R({\bf x},t)e^{i\widehat{\Theta}({\bf x},t)}= 
 \int_{\mathbb{R}^2} G(|{\bf x-x'}|)e^{i\phi({\bf x'},t)}d{\bf x'} \label{eq:R}
\end{equation} 
so that~(\ref{eq:dphi}) can be rewritten as
\begin{equation}
    \frac{\partial \phi({\bf x},t)}{\partial t}=
\omega-R({\bf x},t)\sin{[\phi({\bf x},t)-\widehat{\Theta}({\bf x},t)+\alpha]}. \label{eq:dphiA}
\end{equation}
Now seek statistically steady solutions of Eqs.~(\ref{eq:dphi}), (\ref{eq:G}).
For this special class of solutions we can find a rotating coordinate frame (in state space, not physical space) in which
the phase-locked oscillators in the spiral arms appear motionless, and the mean field becomes stationary across the entire pattern~\cite{lai09B,shikur04}.  To define this rotating frame, let
 $\psi({\bf x},t)\equiv\phi({\bf x},t)-\Omega t$ and 
$\Theta\equiv\widehat{\Theta}-\Omega t$ where $\Omega$ is as yet unknown.  Then 
using~(\ref{eq:R}) and~(\ref{eq:dphiA})
we find
\begin{equation}
    \frac{\partial \psi({\bf x},t)}{\partial t}=
\omega-\Omega-R({\bf x})\sin{[\psi({\bf x},t)-\Theta({\bf x})+\alpha]} \label{eq:dpsi}
\end{equation}
where the time-independent mean field is given by 
\begin{equation}
    R({\bf x})e^{i\Theta({\bf x})}
=\int_{\mathbb{R}^2} G(|{\bf x-x'}|)e^{i\psi({\bf x'},t)}d{\bf x'}. \label{eq:R1}
\end{equation} 
Oscillators for which $|\omega-\Omega|<R({\bf x})$ will reach a steady state, $\psi^*({\bf x})$,
for which $\partial \psi/\partial t=0$ in Eq.~(\ref{eq:dpsi}), whereas those for which 
$|\omega-\Omega|>R({\bf x})$ will drift but distribute themselves with a stationary density in the $\psi$ variable inversely proportional to their angular velocity $\partial \psi/\partial t$~\cite{shikur04,abrstr06}. Equation~(\ref{eq:R1}) can then be solved 
self-consistently by replacing $e^{i\psi}$ by
$e^{i\psi^*}$ (for locked oscillators) and by the mean of $e^{i\psi}$ (for drifting oscillators), where the mean is  
calculated using the density just mentioned.
The result is the nonlinear integral equation
\begin{eqnarray}\label{eq:2Dintegraleqn}
    R({\bf x})e^{i\Theta({\bf x})} & = &
ie^{-i\alpha}\int_{\mathbb{R}^2} G(|{\bf x-x'}|)e^{i\Theta({\bf x'})} \nonumber \\
 & \times & \left(\frac{\Delta-\sqrt{\Delta^2-R^2({\bf x'})}}{R({\bf x'})}\right)d{\bf x'} \label{eq:R2}
\end{eqnarray} 
where $\Delta\equiv \omega-\Omega$.  Figures~\ref{fig:spi1}(c) and \ref{fig:spi1}(d) show $\sin{\Theta({\bf x})}$ and $R({\bf x})$ for a typical spiral wave chimera. 

Our new results concern the analytical solution of Eq.~(\ref{eq:R2}).  
The first step is to simplify it using the spiral wave ansatz of Cohen et al.~\cite{cohneu78}.  Considering
the appearance of the functions $\Theta$ and $R$ in Figs.~\ref{fig:spi1}(c) and (d), 
it is natural to
use the ansatz $R({\bf x})=A(r)$ and
$\Theta({\bf x})=\theta+\Psi(r)$, where $(r,\theta)$ are polar coordinates. Substituting
this into~(\ref{eq:R2}) 
one obtains
\begin{eqnarray}
   A(r)e^{i\Psi(r)} & = & ie^{-i\alpha}\int_0^{\infty}K(r,s)e^{i\Psi(s)} \nonumber \\
   & \times & \left(\frac{\Delta-\sqrt{\Delta^2-A^2(s)}}{A(s)}\right)ds \label{eq:reduced}
\end{eqnarray}
where
\begin{eqnarray}
   K(r,s) & = & 2s\int_0^{\pi}G\left(\sqrt{r^2+s^2-2rs\cos{\theta}}\right)\cos{\theta}\ d\theta \nonumber \\
   & = & 2se^{-(r^2+s^2)}I_1(2rs) \label{eq:K}
\end{eqnarray}
and $I_1$ is the modified Bessel function of the first kind of order $1$. 
This ansatz
has replaced the problem of finding the functions of two variables, $R({\bf x})$ and
$\Theta({\bf x})$, by that of finding $A(r)$ and $\Psi(r)$, both functions of a single variable.

Nevertheless, solving Eq.~(\ref{eq:reduced}) is still a formidable task.  To make progress, we use perturbation theory to analyze~(\ref{eq:reduced}) in the limit of small $\alpha$.
First observe that at $\alpha=0$, Eq.~(\ref{eq:dphi}) is a gradient system in a coordinate frame rotating at speed $\omega$; hence, in this frame all the attractors for~(\ref{eq:dphi}) must be fixed points.  Back in the original frame, these fixed points correspond to phase-locked states rotating at frequency $\Omega = \omega$.  Therefore $\Delta=0$ when $\alpha=0$~\cite{abrstr06}, which motivates the series expansions 
\begin{subequations}
\begin{eqnarray}
   \Delta & = & \Delta_1\alpha+O(\alpha^2) \label{eq:series1} \\
   A(r) & = & A_0(r)+A_1(r)\alpha+O(\alpha^2) \\
   \Psi(r) & = & \Psi_0(r)+\Psi_1(r)\alpha+O(\alpha^2) \label{eq:series3}
\end{eqnarray}
\end{subequations}
as $\alpha \rightarrow 0$.  Substituting~(\ref{eq:series1})-(\ref{eq:series3}) 
into~(\ref{eq:reduced}) we obtain, to lowest order,
\begin{equation}
   A_0(r)[1+i\Psi_0(r)]=\int_0^\infty K(r,s)[1+i\Psi_0(s)]\ ds.
\end{equation}
Hence
\begin{eqnarray}
   A_0(r) & = & \int_0^\infty K(r,s)\ ds \nonumber \\
   & = & (\sqrt{\pi}/2)re^{-r^2/2}\left[I_0(r^2/2)+I_1(r^2/2)\right] \label{eq:A0}
\end{eqnarray}
and
\begin{equation}
   A_0(r)\Psi_0(r)=\int_0^\infty K(r,s)\Psi_0(s)\ ds. \label{eq:psi0}
\end{equation}
Equation~(\ref{eq:A0}) shows that at leading order, the amplitude $A_0(r)$ of the mean field increases with $r$  and satisfies $A_0(0)=0$ and $\lim_{r\rightarrow\infty}A_0(r)=1$, in agreement with the
behavior seen when $\alpha\neq 0$ (Fig.~\ref{fig:spi1}(d)).  Furthermore, under the physically reasonable assumption that $\Psi_0(r)$ can be written as a power series in $r$, one
can show that
$\Psi_0(r)=C$ is the only solution of~(\ref{eq:psi0});
then, from the rotational invariance of the problem we can set $\Psi_0(r)=0$.  This result indicates that the spiral arms approach radial straight lines as $\alpha \rightarrow 0$, a fact we have also confirmed numerically.

At $O(\alpha)$
we obtain $A_1(r)=0$ and
\[
   A_0(r)\Psi_1(r) = \int_0^{\infty}K(r,s)\left[\Psi_1(s)+\frac{\Delta_1}{A_0(s)}\right]ds-A_0(r)
\]
which, after defining $f(r)=A_0(r)\Psi_1(r)$, yields an
inhomogeneous Fredholm equation of the second kind:
\begin{equation}
    f(r)-\int_0^\infty\frac{K(r,s)}{A_0(s)}f(s)ds=\Delta_1\int_0^\infty\frac{K(r,s)}{A_0(s)}ds-A_0(r).
\label{eq:f}
\end{equation}
We have been unable to solve~(\ref{eq:f}) analytically, but extensive numerical investigations 
suggest that to satisfy $f(0)=0$ (which we know to be true, given that $A_0(0)=0$), $\Delta_1$
must be very close to 1. To show that this is plausible we note that for $r^* \gg 1$, 
\begin{equation}\label{eq:approximatesol}
   A_0(r^*)\approx 1 \hspace{5mm} \mbox{ and } \hspace{5mm} K(r^*,s)\approx\frac{e^{-(r^*-s)^2}}{\sqrt{\pi}},
\end{equation}
so that if $f(r)$ is slowly varying near $r^*$, the left side of~(\ref{eq:f}) is approximately zero while
the right side of~(\ref{eq:f}) is approximately $\Delta_1-1$. 

\begin{figure}
	\includegraphics[width=3.2in]{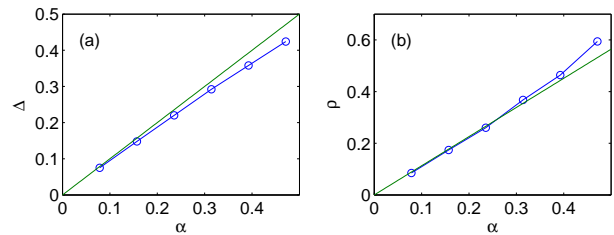}
	\caption{\label{fig:pred} (Color online) Comparison of results from simulation and perturbation theory. 	(a):   Values of $\Delta$ from simulation (circles)	
	compared with predicted values $\Delta=\alpha$ (solid line). 
	(b): Comparison of measured values  of $\rho$ (circles) 
	with $\rho=(2/\sqrt{\pi})\alpha$ (solid line).
	The simulation was run on a square domain of side length $10$, 
	with $201\times 201$ oscillators.}
\end{figure}

Recalling that it is those oscillators
with $A(r)< |\Delta|$ which drift, we can see that to order $\alpha$ the radius of the incoherent
core is $\rho$, where $A_0(\rho)=|\Delta_1| \alpha$. In the limit as $\alpha\rightarrow 0$,
$A_0(\rho)\rightarrow A_0'(0)\rho=(\sqrt{\pi}/2)\rho$, and hence, using $\Delta_1=1$, we have
\begin{equation}\label{eq:radius}
   \rho=\left(\frac{2}{\sqrt{\pi}}\right)\alpha+O(\alpha^2).
\end{equation}
Thus, to order $\alpha$ we have two testable predictions regarding the behavior of the spiral wave chimera as 
$\alpha\rightarrow 0$:
\begin{enumerate}
\item 
Spiral arms 
rotate at angular velocity $\Omega = \omega - \alpha$.
\item 

Incoherent core radius is given by $\rho=(2/\sqrt{\pi})\alpha$.
\end{enumerate}
Figure~\ref{fig:pred} shows measurements of both $\Delta \equiv \omega - \Omega$ and the radius of the incoherent core from a simulation
of Eq.~(\ref{eq:dphi}), as $\alpha$ is varied.
The data points agree very well with the above predictions, despite the use of a finite domain.

\begin{figure}
	\includegraphics[width=3.2in]{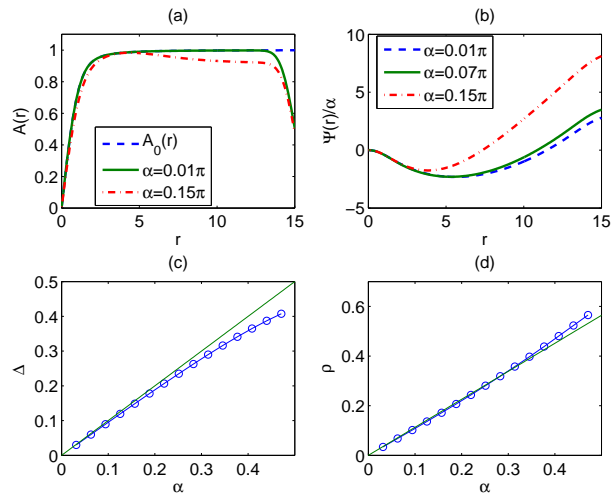}
	\caption{\label{fig:reduced} (Color online) Numerical solutions of Eq.~(\ref{eq:reduced}). (a): $A(r)$
	for two different values of $\alpha$, and $A_0(r)$.
	(b): $\Psi(r)/\alpha$ for three different values of $\alpha$.
	(c): Calculated values of $\Delta$ (circles)
	, compared with the predicted result of $\Delta=\alpha$ (solid line).
	(d): 
	Radius of incoherent core $\rho$: solutions of $A(\rho)=\Delta$ (circles)
	, compared with the predicted value
	$\rho=(2/\sqrt{\pi})\alpha$ (solid line).}
\end{figure}

Along with testing our analytical results against simulations, we have also tested them against numerical solutions of the integral equation~(\ref{eq:reduced}).
The results for a domain $0\leq r\leq 15$ are shown in Fig.~\ref{fig:reduced} as $\alpha$ is varied. 
Figure~\ref{fig:reduced}(a)
compares $A_0(r)$, the only non-trivial zeroth order term in our series expansion, 
to the computed $A(r)$ obtained for
two different values of $\alpha$. Boundary effects near $r=15$ are clearly visible.
For $\alpha=0.01\pi$, $A(r)$ is indistinguishable from $A_0(r)$
for $r< 13$. Figure~\ref{fig:reduced}(b) plots $\Psi(r)/\alpha$ for three different values of $\alpha$; this choice of dependent variable is motivated by Eq.~(\ref{eq:series3}), which predicts that $\Psi(r)$ should  be proportional to $\alpha$, for $\alpha$ small.
Figures~\ref{fig:reduced}(c) and \ref{fig:reduced}(d) show $\Delta$ and $\rho$ as functions of $\alpha$, now found by
solving~(\ref{eq:reduced}) instead of simulating~(\ref{eq:dphi}), in the same format as
Fig.~\ref{fig:pred}. The agreement with the data in Fig.~\ref{fig:pred} is excellent.

\begin{figure}
	\includegraphics[width=3.2in]{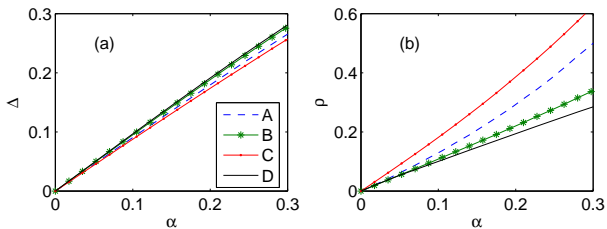}
	\caption{\label{fig:kern} (Color online) Effects of different kernels.  Panels (a) and (b) show $\Delta$ and $\rho$ respectively
for kernels $2\pi G(r)$ given by: (A)~$ K_0(r)$; (B)~$2e^{-r^2}$; (C)~$e^{-r}$; (D)~$2H(r)H(1-r)$, where $H$ is the
Heaviside step function.}
\end{figure}

The results above were derived using the Gaussian kernel~(\ref{eq:G}).  In contrast, Shima and Kuramoto~\cite{shikur04} used the modified Bessel function of the second kind, $K_0$, as a kernel. This coupling function arises naturally in a class of reaction-diffusion systems where active chemical elements are indirectly coupled by passive, fast diffusive agents~\cite{shikur04}.  Using a Gaussian allowed us to explicitly perform the integral in~(\ref{eq:K})
and thus make analytical progress in the subsequent perturbation analysis.  
For $K_0$ and other kernels one can evaluate the integral in~(\ref{eq:K}) numerically. In Fig.~\ref{fig:kern}
we show $\Delta$ and $\rho$ as a function of $\alpha$ for four different kernels, found by
numerically solving~(\ref{eq:reduced}). Interestingly, we see that while $\Delta$ does not seem to depend on the
details of the kernel, $\rho$ does.

Several puzzles remain about the spiral wave chimera~\cite{kurshi06}.  For example, we do not understand how it bifurcates.  When $\alpha$ is increased sufficiently far from zero, the chimera becomes numerically
unobservable.  Perhaps, for example, it loses stability though a subcritical Hopf bifurcation, 
or is destroyed through a 
saddle-node bifurcation.  Even when the chimera does exist, its core wanders
erratically. This can be rationalized by observing that there is no preferred origin
for the spiral in an infinite domain, so it is neutrally stable with respect to translations. 
The finite number of incoherent
oscillators act a source of effectively random ``noise'' which then jiggles the
spiral in these neutral directions.

A second puzzle is that the spiral wave chimera apparently exists only if $\alpha$ is sufficiently close to 0, whereas the chimeras found in lower-dimensional systems~\cite{abrstr04,abrstr06, abrmir08,pikovskyPRL08,lai09A} exist only if $\alpha$ is sufficiently close to $\pi/2$.  In physical terms, two-dimensional chimeras live near the gradient system limit, whereas one- and zero-dimensional chimeras live in the opposite regime, where the dynamics have an almost reversible or conservative character.  At the moment we have no explanation for this difference.

Although we have assumed identical oscillators in our analysis, this is not necessary.  Spiral wave chimeras should persist in the presence of slight disorder.  
For example, \citet{lai09B} studied a variant of Eq.~(\ref{eq:dphi}) for non-identical oscillators and with a
kernel $G$ which was zero for $|{\bf x-x'}|$ larger than a particular radius. He used the recent
ansatz of Ott and Antonsen~\cite{ottant08, ottant09} to derive a differential equation
for a variable closely related to the order parameter.  
For oscillators whose intrinsic frequencies $\omega$
are randomly drawn from a Lorentzian distribution, the steady states 
of this differential equation satisfy~(\ref{eq:R2}), where $\Delta=\omega-\Omega+iD$ and $D$
is the half-width-at-half-maximum of the Lorentzian distribution. The results 
in Fig.~\ref{fig:pred} agree qualitatively with those of~\citet{lai09B}. 

The possibility of observing spiral wave chimeras in physical systems naturally arises. Nonlocal
coupling of the form used here occurs in some networks of neurons~\cite{laicho01,horoph97}, so such systems may be the best candidates. We leave the experimental observation of chimera states as a
challenge to others.

Research supported in part by NSF grant CCF-0835706. We thank S. Shima for helpful advice about the simulations.


\end{document}